\documentclass[11pt,dvips,psfig]{article}
\newlength{\bredde}
\def\slash#1{\settowidth{\bredde}{$#1$}\ifmmode\,\raisebox{.15ex}{/}
\hspace*{-\bredde} #1\else$\,\raisebox{.15ex}{/}\hspace*{-\bredde} #1$\fi}
\newcommand{\be}{\begin{equation}}
\newcommand{\ee}{\end{equation}}
\newcommand{\ba}{\begin{array}{ccc}}
\newcommand{\ea}{\end{array}}

\newcommand{\noi}{\vspace{12pt}\noindent}
\newcommand{\lG}{\raise.3ex\hbox{$\stackrel{\leftarrow}{G}$}}
\newcommand{\lU}{\raise.3ex\hbox{$\stackrel{\leftarrow}{U}$}}
\newcommand{\lP}{\raise.3ex\hbox{$\stackrel{\leftarrow}{{\cal P}}$}}
\newcommand{\leta}{\raise.3ex\hbox{$\stackrel{\leftarrow}{\eta}$}}
\newcommand{\lOmega}{\raise.3ex\hbox{$\stackrel{\leftarrow}{\Omega}$}}
\newcommand{\ldr}{\raise.3ex\hbox{$\stackrel{\leftarrow}{\delta^r}$}}

\def\log{\textnormal{log}}
\def\det{\textnormal{det}}
\def\exp{\textnormal{exp}}
\def\bea{\begin{eqnarray}}
\def\eea{\end{eqnarray}}

\def\gtwid{\raise.3ex\hbox{$>$\kern-.75em\lower1ex\hbox{$\sim$}}}
\def\ltwid{\raise.3ex\hbox{$<$\kern-.75em\lower1ex\hbox{$\sim$}}}

\def\r{\ref}
\def\al{\alpha}
\def\de{\delta}
\def\la{\lambda}

\def\om{\omega}
\def\a{\alpha}

\def\m{\mu}

\begin{document}

\topmargin -1.4cm
\oddsidemargin -0.8cm
\evensidemargin -0.8cm
\title{\Large{{\bf The zeros of the QCD partition function} \\ 
}}

\vspace{1.5cm}
\author{
{\sc A.~D. Jackson}$^a$,
{\sc C.~B. Lang}$^b$,
{\sc M. Oswald}$^a$,\\
and {\sc K. Splittorff}$^a$ \\ \\
$^a$ The Niels Bohr Institute\\ Blegdamsvej 17\\ DK-2100 Copenhagen {\O},
Denmark\\ \\
$^b$ Institut f\"ur Theoretische Physik,\\
Universit\"at Graz, A-8010 Graz, Austria }
\date{\today} 
\maketitle
\vfill
\begin{abstract} 
We establish a relationship between the zeros of the partition function 
in the complex mass plane and the spectral properties of the Dirac operator 
in QCD. This relation is derived within the context of chiral Random Matrix 
Theory and applies to QCD when chiral symmetry is spontaneously broken.  
Further, we introduce and examine the concept of normal modes in chiral 
spectra. Using this formalism we study the consequences of a finite Thouless 
energy for the zeros of the partition function. This leads to the demonstration 
that certain features of the QCD partition function are universal. 
\end{abstract}
\vfill

\thispagestyle{empty}
\newpage
\section{Introduction}

In recent years the spectral correlators of the Dirac operator in QCD
have been the object of intense study using both numerical and analytic 
means.  These correlators contain valuable information regarding both 
the chiral properties of the QCD vacuum and the topological structure of 
the gauge fields.  The relation to the chiral properties of the QCD vacuum 
was established by Banks and Casher \cite{BC}:  The eigenvalue density of 
the QCD Dirac operator at eigenvalue zero is proportional to the chiral 
condensate and is therefore an appropriate order parameter for chiral 
symmetry.  As a complement to the Banks-Casher relation, one has the 
Yang-Lee picture \cite{YL52} of a phase transition. In the attempt to 
analyze phase transitions in statistical spin models
Lee and Yang \cite{YL52} introduced the concept
of the zeros of the finite volume partition function in the thermodynamic
limit. The volume dependence of these zeros allows finite size
scaling studies and subsequent identification of universality classes. 
In the case of chiral symmetry, one focuses on the zeros of the partition 
function  in the complex quark mass plane.  
If these zeros pinch the real axis and exhibit a constant density, a 
discontinuity in the partition function arises at the pinch, and chiral 
symmetry is spontaneously broken.  The mass zeros of the partition function 
and the low lying eigenvalues of the Dirac operator thus contain similar 
information about the chiral phase transition.  The relation between the 
two is, however, involved.  The partition function and its zeros are obtained 
by averaging over all gauge field configurations. By contrast, the eigenvalues 
of the Dirac operator are given for each gauge field configuration, and only 
a density of eigenvalues is well-defined after averaging.

The challenge of deriving relations between the zeros of the partition 
function and the eigenvalues of the Dirac operator was first taken up 
by Leutwyler and Smilga \cite{LS}.  They studied QCD in a Euclidean 
4-dimensional box with side length $L$ subject to the constraint that 
\be
\frac{1}{\Lambda} \ll L \ll \frac{1}{m_{\pi}} \ .
\label{RMTregime}
\ee
Here $m_{\pi}$ is the pion mass and $\Lambda$ is the typical QCD scale.
They computed the QCD partition function for equal quark masses using the
effective chiral Lagrangian and found that quark masses enter only in the
rescaled combination $m\,L^4\,\Sigma$, where $\Sigma$ is the chiral condensatein the chiral limit.  They further observed that 
the partition function zeros could be thought of as average positions of the
eigenvalues.  While highly suggestive, these results were not completely 
quantitative.  The situation has changed dramatically since then. The main
break through came with the introduction \cite{SV} of random matrix concepts 
in QCD which permits the study of the correlations of the eigenvalues of 
matrices drawn on a general weight constructed to ensure the chiral structure 
of each eigenvalue spectrum.  The relation of random matrix theory 
to QCD in the limit (\ref{RMTregime}) has been established 
through a number of universality studies \cite{RMTuni}, lattice QCD
simulations \cite{Latticebulk},\cite{Latticemicro}, and direct calculations 
using the effective 
chiral Lagrangian \cite{CPT}.  (For a review of random matrix theory in QCD 
see \cite{VW}.)  In terms of the spectral correlation functions, the universal 
limit in which QCD and chiral random matrix theory ($\chi$RMT) coincide is 
the limit 
\be
N\to\infty\;,\quad \lambda\to 0\;,\quad  m \to 0\;,
\end{equation}
in which the microscopic variables
\begin{equation}
\zeta\equiv  2\,N\,\lambda\;,\quad\mu\equiv  2\,N \,m
\ee
are kept fixed and $N$ is identified as the dimensionless volume.  (Here, $\lambda$ denotes an eigenvalue of the Dirac operator 
and $m$ is the  dimensionless quarkmass parameter.)  The determination of the individual eigenvalue 
distributions and their most important correlators now permits direct 
comparison of partition function zeros and eigenvalue positions.  The 
suggestion of Leutwyler and Smilga is remarkably accurate.  The zeros and 
the average positions of the eigenvalues are intimately connected.  

Here we shall demonstrate that this relationship can be understood as a 
fundamental property of the chiral ensembles.  We show that the zeros are
uniquely trapped by the maxima of the joint eigenvalue distribution function.
This trapping appears on all scales and is thus relevant for any finite $N$ 
as well as in the large $N$ limit.  To obtain a better understanding of the 
relation between the maxima of the joint eigenvalue distribution function and 
the zeros of the partition function, we introduce and determine the spectral 
``normal modes'' of the chiral unitary ensemble.  This provides us with a 
simple tool to describe the fluctuations of the eigenvalues about the maximum 
of the joint eigenvalue distribution.  

As suggested in (\ref{RMTregime}), chiral random matrix theory is not 
expected to describe all aspects of QCD.  Only correlations below a certain
energy length are expected to be in agreement with chiral random matrix
theory \cite{Thouless}. In solid state physics, this energy is denoted as the
Thouless energy.  Recently \cite{JMRSW} it was realized that the effects of 
a finite Thouless energy can be studied naturally using the language of 
spectral normal modes \cite{AJP}.  We thus perform a normal mode analyses 
of the chiral ensemble to formulate and establish certain universal 
features of the partition function zeros.  This argument is independent of 
standard proofs of universality, and its general nature can shed some 
light on the way universality is realized.

In section \ref{sect:chiRMT} we show that for $N_f=1$ the zeros of the
partition function of $\chi$RMT are trapped between the maximum positions
of the joint distribution function. This result holds for all scales. 
In section \ref{sect:chiNM} we derive the normal modes of the chiral
unitary ensembles and find that they are Chebyshev polynomials in the 
large $N$ limit.  We discuss the effects of the Thouless energy in section
\ref{sect:Thouless}.  In section \ref{sect:micro} we make the connection
to the familiar microscopic spectral density.  Our conclusions are contained 
in section \ref{sect:conc}.

\section{Zeros of the partition function in $\chi$RMT}
\label{sect:chiRMT}

\subsection{Chiral random matrix theory}

We start with the partition function of chiral random matrix theory
($\chi$RMT) for $N_f$ flavors, which is given by \cite{SV, HV}
\be
Z_{N, \beta}^{N_{f},\nu}(\{m_{f}\}) = \int{}DW
\prod\limits_{f=1}^{N_{f}} \det(D + m_{f})\;
\exp\left[-\frac{N\,\beta\,\Sigma^2}{2}\,{\rm Tr} (W^{\dag}W)\right]\;,
\ee
where $\beta$ denotes the Dyson index and $DW$ is the Haar measure over the
Gaussian distributed random matrices $W$. $D$ is the analogue of the 
Dirac operator which has the chiral structure
\be
D = \left( \begin{array}{cc} 0 & iW \\ iW^{\dagger} & 0 \end{array}
\right)\;.
\ee
Here $W$ is a $N\times M$ matrix with $\nu = |N-M|$ playing the r\^ole of
the topological charge. Without loss of generality we assume $\nu$ to be 
positive. The chiral condensate in the chiral limit, $\Sigma$, is related 
to the eigenvalue density of $D$, $\rho(\la)$, via the Banks--Casher relation
\cite{BC}
\be
\Sigma = \lim_{\lambda\to 0}\, \lim_{m_f\to 0}\,\lim_{N\to\infty} 
\frac{\pi\,\rho(\lambda)}{N}\;.
\ee
The partition function is invariant under transformations
$W$ $\rightarrow$ $U^{\dag}\,W\,V$,  where $U$ is a $N \times N$ matrix 
and $V$ a $M \times M$ matrix.  Following the diagonalization 
$W = U^{\dag}\Lambda{}V$, the partition function can be expressed in terms 
of the eigenvalues of $W$ (with $\Sigma\equiv 1$),
\be\label{pf}
Z_{N, \beta}^{N_{f},\nu}(\{m_{f}\})  =  
\left(\prod_{f=1}^{N_f} \,m_{f}^{|\nu|}\right)
\int_{-\infty}^{+\infty}\ldots\int_{-\infty}^{+\infty}
\prod\limits_{k=1}^{N}\left[d\lambda_{k}\prod_{f=1}^{N_f}\big(\lambda_k^2
+ m_f^2\big)\lambda_k^{\beta\nu + \beta - 1}
e^{-\frac{N\beta}{2}\lambda_k^2}\right]\Delta\big( \lambda^{2}
\big)^{\beta} .
\ee
The Vandermonde determinant, $\Delta\big( \lambda^{2} \big)$, which is 
the non-trivial Jacobian of the transformation from the matrices to the
eigenvalues, has the form 
\be
\Delta\big( \lambda^{2} \big) = \prod\limits_{k<l}\big(\lambda_{k}^{2} -
\lambda_{l}^{2} \big)~.
\ee
The partition function (\ref{pf}) can now be written as an integral over
the joint probability density $P_{N,\beta}^{N_{f},
\nu}(\lambda_1,\ldots,\lambda_N; \{m_f\})$ as
\be
Z_{N, \beta}^{N_{f},\nu}(\{m_{f}\}) =
\int_{-\infty}^{+\infty}\ldots\int_{-\infty}^{+\infty}\prod\limits_{k=1}^{N}d\lambda_{k}
P_{N,\beta}^{N_{f}, \nu}(\lambda_1,\ldots,\lambda_n; \{m_f\})
\ee
with 
\be\label{P}
P_{N,\beta}^{N_{f}, \nu}(\lambda_1,\ldots,\lambda_N; \{m_f\}) =
\left(\prod_{f=1}^{N_f} m_{f}^{|\nu|}\right) \prod_{k=1}^{N}
\left[\prod_{f=1}^{N_f}\big(\lambda_k^2 +
m_f^2\big)\lambda_k^{\beta\nu + \beta - 1}
e^{-\frac{N\beta}{2}\lambda_k^2}\right]\,\Delta\big( \lambda^{2}
\big)^{\beta}\;.
\ee
Unlike real QCD, $\chi$RMT has the special feature that the partition 
function can be expressed in terms of the eigenvalues of the Dirac operator. 
This enables us to derive a number of statements regarding the zeros of the 
partition function.  We now focus on the case $\beta=2$ --- the universality 
class of QCD with 3 colours and quarks in the fundamental representation of 
the gauge group.  (The choice $\beta=1$ corresponds to QCD with two colours 
in the fundamental representation; $\beta=4$ describes QCD with any number 
of flavours and quarks in the adjoint representation of the gauge group.)

Eq.\,(\ref{P}) expresses an evident duality between flavor and topology: 
The joint probability density for $N_v$ massless flavours and $N_f$ massive 
flavours depends only on $\nu+N_v$.  This relation was proven for the QCD 
partition function independent of $\chi$RMT in \cite{DV}.   

We now wish to determine the maximum of the joint probability distribution.  
This will allow us to put a tight bound on the zeros of the partition 
function.  The chiral normal modes, to be discussed in section 
\ref{sect:chiNM}, describe fluctuations about the maximum of the joint 
probability distribution. 

\subsection{Extremum of the joint probability distribution}

In order to determine the maximum of the joint eigenvalue probability
distribution, we consider variations of $\log P_{N}^{N_f,\nu}$ with 
respect to the eigenvalues.  (We assume the eigenvalues to be ordered with
$\lambda_{i}<\lambda_{i+1}$.)  We introduce the coordinates $y_{i} = 
\la_{i}^2$ and evaluate the equations 
\be
\frac{\partial\;\log\;P_{N}^{N_f, \nu}}{\partial y_{i}} = 0~.
\ee
For $N_f=0$ and topological sector $\nu$ this yields
\be\label{equ}
\left(\nu+\frac{1}{2}\right)\frac{1}{N\,y_i} - 1 + \frac{1}{N} \, 
\sum_{j\neq i}\frac{2}{y_i - y_j} = 0\;.
\ee
We now choose to focus on the quenched, i.e. $N_f=0$, joint eigenvalue 
probability distribution. The solution to this equation reveals that 
the maximum of $\log P_{N}^{N_f=0,\nu}$ is obtained for 
\be
L_{N}^{\nu-1/2}\left(N\lambda_i^2\right) = 0 \;,
\label{max}
\ee
where $L^\al_N$ denotes the generalized Laguerre polynomials.
This result follows from the observation that Laguerre's differential 
equation,
\be
z L_{N}^{\a}(z)'' + (\a +1-z)L_{N}^{\a}(z)' + N L_{N}^{\a}(z)=0~,
\ee
reduces to (\ref{equ}) at the zeros of $L_{N}^{\nu-1/2}$. (The proof follows
from considerations similar to those made in appendix A6 in \cite{Mehta}).
For $\nu=0$ we can use the fact that
\be
L_n^{-1/2}(z^2) = \frac{(-1)^n}{n!\,2^{2n}}
H_{2n}(z)
\ee
to see that the density of eigenvalues in the $N\to\infty$ limit is precisely that of the usual
Gaussian ensembles, i.e. a semicircle with support $-2 \le \lambda \le +2$.  
The partition function for $\beta=2$ is the average of a product of
fermionic determinants over $P_{N}^{N_f=0,\nu}$
\be
Z^{N_{f},\nu}_{N}(\{m_{f}\}) =
\left\langle
\prod_{f=1}^{N_{f}}\,m_{f}^{|\nu|}\prod_{i=1}^{N}(\lambda_i^2+m_{f}^2)
\right\rangle\;.
\ee
This can be readily evaluated using orthogonal polynomials, and the result
for $\nu=0$ agrees with the one presented in \cite{WGW}
\be
Z^{N_{f},\nu=0}_{N}(\{m_{f}\}) =
\prod_{f=1}^{N_{f}}\frac{(N+f-1)!}{N^{N+f-1}}
\frac{C_N^{N_{f}}\left(\{m_{f}\}\right)}{\Delta_{N_{f}}\left(\{-m_{f}^2\}\right)}~, 
\ee
where 
\be
C_N^{N_{f}}=\det\left[L_{N+f-1}^{(0)}\left(-N
m_{f'}^2\right) \right]_{f,f'=1,\ldots,N_{f}}~,
\ee
and $\Delta_{N_{f}}\left(\{-m_{f}^2\}\right)$ is the Vandermonde
determinant with the negative square of the $N_f$ masses as arguments. For
the special case of $N_{f}=1$ this yields the result
\be\label{nf1}
Z^{N_f=1,\nu=0}_{N}(m) = \frac{N!}{N^N}L_{N}^{(0)}\left(-N m^2\right)~,
\ee
which is now ready for investigation. The expression (\ref{nf1}) coincides 
(up to a constant) with the series expansion for the partition
function derived in \cite{HJV}.  From the expression found there we find
that  
\be
Z^{N_f=1,\nu}_{N}(m) \sim m^{|\nu|} L_{N}^{(\nu)}\left(-N m^2\right)~.
\ee

\subsection{Trapping of the zeros}
\label{sect:trap}

The closed forms given above allow us obtain information about the zeros of 
the partition from the spectral correlators.  Specifically, we now
show that the locations of the partition function zeros in $\chi$RMT are
trapped by the maxima of the joint distribution function. 

The Laguerre polynomials $L_n^\alpha(z)$ are polynomials orthogonal on the 
interval $[0,\infty ]$ with weight function $w(z)=z^\alpha\exp(-z)$.  They 
have three properties which are useful for our purpose: 
\begin{itemize}
\item For orthogonal polynomials in general, the zeros of the $N-$th order
polynomial and the end-points of the weight function define $N+1$ intervals.  
Exactly one zero of the orthogonal polynomial of order $(N+1)$ lies in each 
of these intervals.
\item  For fixed $N$, the $i$th zero of the Laguerre polynomial,
$L_{N}^a(x)=0$, is a monotonically increasing function of $a$, thus
\be
\label{B2}
\frac{dx_{i}}{da} > 0\;.
\ee
\item The generalized Laguerre polynomials are related to one another 
via
\be
L_N^{a+1}(x) = - \frac{d L_{N+1}^a(x)}{dx}~.
\label{B3}
\ee
\end{itemize}
In the last section we saw that the massless joint distribution function
has its maxima when the eigenvalues are located at the zeros of 
$L_{N}^{-1/2}(N\lambda^2)$ and that the partition function for $N_f=1$ 
and $\nu=0$ is proportional to $L_{N}^{0}(-N m^2)$.  We are interested in 
relating the zeros of the partition function to the position of the 
eigenvalues at the maximum of the joint distribution function.   Since 
the zeros of $L_{N}^{0}(-N\,m^2)$ follow from those of $L_{N}^{0}(N\,m^2)$ 
by a rotation from the real to the imaginary axes in $m$, we restrict our 
attention in the following considerations to Laguerre polynomials of 
positive argument.

It follows from (\ref{B2}) that the zeros of $Z^{N_f=1,\nu=0}_{N}$ are 
trapped between the corresponding zeros of $L_{N}^{-1/2}$ and 
$L_{N}^{+1/2}$.  According to (\ref{B3}), the zeros of $L_N^{1/2}$ lie at 
the extrema of $L_{N+1}^{-1/2}$.  The extrema of $L_{N+1}^{-1/2}$ are evidently 
trapped by the zeros of $L_N^{-1/2}$ (and the end points, if necessary).  
The result is that the $i$th zero of $L_N^{0}$ is trapped between the 
$i$th and $(i+1)$st zero of $L_N^{-1/2}$ and, given the nature of the 
argument, is expected to be closer to the lower value.  In other words, 
the zeros of $Z^{N_f=1,\nu=0}_N(m)$ are trapped by the most probable values for the 
eigenvalues of $P_{N}^{N_{f}=0,\nu=0}$.  This result is an exact property 
of the chiral unitary ensembles and is consequently valid on every scale 
including the microscopic scale.  

The trapping just derived relates the zeros of the $N_f=1$ partition
function and the location of the maximum in the joint eigenvalue distribution 
of the quenched $(N_f=0)$ ensemble.  We can also relate the zeros of the 
$N_f=1$ partition function to the maximum of its own integration kernel, 
$P^{N_f=1,\nu}_N(\{\lambda_i\},m)$.  For $m \to 0$ we have
$P^{N_f=1,\nu}_{N}(\{\lambda_i\},m)=P^{N_f=0,\nu+1}_{N}(\{\lambda_i\})$ by
flavor-topology duality.  The other limit, $m\to\infty$, decouples one
flavor and leaves $P_{N}^{N_{f}=0,\nu}$.  So, with increasing $m$, the
extrema of $P_{N}^{N_{f}=1,\nu}(m)$  move smoothly from those of 
$P_{N}^{N_{f}=0,\nu+1}$ to those of $P_{N}^{N_{f}=0,\nu}$, i.e. 
from the zeros of $L_{N}^{\nu-1/2}$ to the zeros of $L_{N}^{\nu+1/2}$.  
By (\ref{B2}), they must pass the zeros of the partition function 
$L_{N}^{\nu}$ for some intermediates values of the mass $m$.  The relation 
between the collective maximum of $\log P$ and the average eigenvalue 
positions will be reconsidered in section \ref{sect:micro}.

\section{Normal modes in $\chi$RMT}
\label{sect:chiNM}
 
\noi
We have seen that the maximum of the massless joint distribution
function is obtained when the eigenvalues are located at the zeros of 
the Laguerre polynomials.  In order to study the properties of 
fluctuations about this maximum, it is useful to make a Gaussian 
approximation to $P_{N}^{N_f=0,\nu}$ which leads to the form
\be\label{log}
\log\;P_{N}^{N_f=0,\nu} \approx \log\;P_{N;(0)}^{N_f=0,\nu} +
\frac{1}{2}\delta\lambda_{i}\,C_{ij}\,\delta\lambda_{j}~,
\ee
where $\delta\lambda_{i}$ is the position of the $i$th eigenvalue relative
to $\bar\la_{i}$, its value at the collective maximum of $\log \;P_{N}^{N_f=0,\nu}$.  The matrix $C$ is defined as
\be
C_{ij} = \frac{\partial^2}{\partial\la_{i}\partial\la_{j}}\log \; 
P_{N}^{N_f=0,\nu}~,
\ee
evaluated at the maximum.  Concentrating again on the case $N_f=0$, we 
find that the diagonal elements of $C$ are
\be
C_{ii} = -2N - \frac{2\nu+1}{\lambda_{i}^2} - 4\sum_{j\neq
i}\frac{(\lambda_{i}^2+\lambda_{j}^2)}{(\lambda_{i}^2-\lambda_{j}^2)^2}~,
\ee
and that the off-diagonal elements are  
\be\label{C}
C_{ij} = \frac{8\lambda_{i}\lambda_{j}}{(\lambda_{i}^2-\lambda_{j}^2)^2}~.
\ee
We now consider the eigenvalue equation for the real symmetric matrix $C$:
\be
\sum_{i=1}^{N}C_{ij} \phi_{j}^{(k)} = \om_{k} \phi_{i}^{(k)}~,
\ee
The eigenvectors, $\phi^{(k)}$, are the (normalized) normal modes of the 
$\chi$RMT spectrum.  They describe the statistically independent correlated
fluctuations of the eigenvalues of the random matrix about their most 
probable values.  The normal modes provide an alternate description of 
the eigenvalues of any given random matrix since 
\be
\de\la_{i} = \sum_{k=1}^{N} \, c_{k} \phi_{i}^{(k)} \ \ {\rm with}\ \ 
\sum_{i=1}^{N} \, \phi_i^{(k)} \phi_i^{(k')} = \delta_{kk'}~.
\label{deltalambda}
\ee 
We can locate the eigenvalues by specifying either the $\delta 
\lambda_i$ or the amplitudes $c_k$ as convenient.  The eigenvalues, 
$\omega_k$, provide a measure of the magnitude of these fluctuations.  

The derivation of these eigenvalues and eigenvectors can be performed as 
in \cite{AJP}.  The resulting eigenvalues are
\be
\om_{k}= -4\,k\,N~.
\label{freq}
\ee
As in \cite{AJP}, we find a linear dispersion relation valid for all
$k$ and $N$.  The linearity of (\ref{freq}) is a reflection of the well-known 
rigidity of random matrix spectra.  Furthermore, this result is independent 
of $\nu$.  Just as in the Gaussian case, the eigenvectors are found to be 
Chebyshev polynomials in the large $N$ limit (i.e., with corrections of 
order $1/N$):
\be
\label{Cheb}
\phi_{i}^{(k)}=\sqrt{\frac{2}{N}}\;U_{2k-1}\left(\frac{\la_{i}}{2}\right)~.
\ee
The normalization of the eigenvectors is
\be\label{norm}
\int dx\rho(x)\phi^{(k)}(x)\phi^{(l)}(x) = \de_{kl}~,
\ee
where $\rho(x)$ is again the semicircle
\be
\rho(x)=\frac{N}{\pi}\sqrt{4 - x^2}~.
\ee
Note that only odd normal Chebyshev polynomials appear.  This is a 
consequence of chiral symmetry, which ensures that all non-zero 
eigenvalues come in pairs $(-\lambda,\lambda)$.  Equation (\r{log}) can 
now be written as
\be\label{quad}
\log\left(\frac{P_N^{N_f=0,\nu}}{P_{N;(0)}^{N_f=0,\nu}}\right) =
\frac{1}{2}\sum_{k=1}^N|c_{k}|^2\om_{k}~.
\ee
Following (\ref{deltalambda}), the coefficients $c_{k}$ are constructed 
as
\be
c_{k} = \sum_{i=1}^N \, \de\la_{i}\,\phi_{i}^{(k)}\;,
\ee
and are statistically independent.  It is obvious from (\r{quad}) that 
that the mean square amplitude for the $k$th normal mode is 
\be
\langle c_{k}^2\rangle = \frac{1}{|\om_{k}|}~.
\ee
We can use the normal modes to construct a Gaussian approximation to the
partition function as
\be
{\cal Z}_N^{N_f=1,\nu}(m) = \int_0^\infty \prod_{k=1}^N {\rm d}c_k\,\,
\exp{\left[- \frac{1}{2}\sum_{k=1}^N|c_{k}|^2 |\om_{k}| \right]} \,
(m^2+(\bar{\lambda}_k+\delta\lambda_k)^2) \ ,
\label{gaussianZ}
\ee
with the $\bar{\lambda}_k$ given by (\ref{max}) and $\delta\lambda_k$ 
given by (\ref{deltalambda}).  The Gaussian approximation (\ref{quad}) also 
permits a simple approximate calculation of the number variance \cite{AJP}.  
This calculation reveals that the familiar logarithmic behaviour of the 
number variance (i.e., the ``spectral rigidity'' of the random matrix 
ensembles) is a direct consequence of the linearity of the dispersion 
relation (\ref{freq}) for all $k$.  

We have chosen to consider the normal modes for the case $N_f=0$ and 
$\nu=0$.  This choice is some what arbitrary; it would be equally sensible 
to start with the $N_f=0$ and $\nu=1$ normal modes.  We will employ this 
Gaussian approximation below to consider the sensitivity of the partition 
function zeros to the effects of a Thouless energy.  Since the resulting 
shifts are small, this arbitrariness will be of no consequence.  

\section{Effects of a Thouless energy}
\label{sect:Thouless}

Normal modes describe the correlated fluctuations of eigenvalues
about their most probable values.  As we have seen (\ref{freq}), the normal 
modes for $\chi$RMT obey a linear dispersion relation.  By contrast, 
uncorrelated eigenvalues obey a quadratic dispersion relation, and the 
mean square amplitude of the lowest mode with $k=1$ is larger by a 
factor of $N/k$ \cite{AJP}.  In QCD, it is expected that 
spectral correlations in a sufficiently small energy domain will follow 
the predictions of $\chi$RMT.  On larger energy scales, spectral 
correlations die out.  The characteristic energy which divides these 
regions is the Thouless energy, usually denoted by $E_c$.  In 4-dimensional 
QCD the Thouless energy is estimated to be \cite{Thouless} 
\be 
E_c/D\sim \sqrt{N}~, 
\ee 
where $D$ is the mean level spacing. This behaviour has been verified in
lattice studies \cite{ThoulessLattice}.  The connection between the Thouless 
energy and the normal modes of the eigenvalue spectrum has been investigated 
in \cite{JMRSW} for the case of sparse matrices.  There it was found that 
``almost all'' normal modes obey the linear dispersion relation discussed 
above with remarkable accuracy.  Reflecting the presence of a Thouless 
energy, a small fraction (i.e., approximately $1/\sqrt{N}$) of long wave 
length modes are strongly enhanced.  The mean square amplitude of the 
longest wave length mode with $k=1$ approaches the value appropriate for 
uncorrelated eigenvalues.  Since such enhancement can be readily incorporated 
in our Gaussian approximation to the partition function (\ref{gaussianZ}), 
normal modes provide us with a natural and convenient tool to study the 
effects of a Thouless energy on the zeros of the partition function.  So far 
we have seen that the most probable eigenvalues are located at the zeros 
of the Laguerre polynomial $L_N^{-1/2}$ and that the zeros of the partition 
function are given by the zeros of $L_N^{0}$.  The Gaussian approximation 
(\ref{gaussianZ}) allows us to see how this result is modified by the 
presence of a Thouless energy.
To mimic the effects of the Thouless energy, we shall enhance the long
wavelength modes in the partition function.  

Our aim is to demonstrate that the zeros of the partition function in 
the microscopic region remain virtually unaffected  
even if the enhancement of the soft modes is substantial.  Since we are 
concerned only with the microscopic zeros, every long wavelength 
mode contributes to a ``breathing'' of the spectrum.  In order to 
investigate the influence of the soft modes, it is sufficient to evaluate 
the strength (i.e., mean square amplitude) of this effective breathing.  

For concreteness we start with $\sqrt{N}$
longest wavelengths modes with fluctuations as given by the Gaussian
approximation. Let $\bar{\lambda}_i$ denote the values at the maximum of
$P^{N_f=0,\nu=0}_N$. The fluctuations introduced in the Gaussian approximation
by the  $\sqrt{N}$ longest wavelengths modes are
\be
\bar{\la}_i\longrightarrow \bar{\la}_i
\left(1+\sum_{k=1}^{\sqrt{N}}\eta^{(k)}\;c_k\right)
\equiv \bar{\la}_i\,s~.
\ee
{}From (\ref{Cheb}) we know that the normal modes are
$\sqrt{2/N}\;U_{2k-1}\left(\la_{i}/2\right)$. This gives for the linear
coefficient
\be\label{etax}
\eta^{(k)} = \sqrt{\frac{2}{N}}\;k\;(-1)^{k+1}~.
\ee
Additionally we know from (\ref{C}) that the normal modes are linearly
independent. With this information the variance of $s$ becomes in leading 
order of $1/N$
\be\label{s2x}
\langle s^2 \rangle - \langle s \rangle^2 = \langle s^2 \rangle - 1 \sim
\sum_{k=1}^{\sqrt{N}}\;\frac{k^2}{N}\;\frac{1}{k\,N} =
\frac{1}{\,N^2}\sum_{k=1}^{\sqrt{N}} k =
\frac{1}{\,N^2}\sqrt{N}(\sqrt{N}+1) \sim \frac{1}{N}~.
\ee
Here we used the facts that the linear terms vanish and that $\langle
c_{k}^2\rangle = 1/|\om_{k}|$ with (\ref{freq})
for the quadratic terms. The result is simply a $\mathcal{O}(1/N)$ correction.

In order to study the effects of a Thouless energy, we now enhance 
the mean square amplitudes of these $\sqrt{N}$ soft modes 
by a factor of $N/k^2$.  This factor provides a smooth interpolation 
from the behaviour of uncorrelated soft modes (for $k=1$) to that of 
$\chi$RMT (for $k = \sqrt{N}$).  This interpolation is completely 
consistent with the results of \cite{JMRSW}.  We now find that 
\be\label{s2d}
\langle s^2 \rangle - \langle s \rangle^2\sim
\sum_{k=1}^{\sqrt{N}}\frac{k^2}{N} \, \frac{1}{kN} \, \frac{N}{k^2}\sim 
\frac{\log(N)}{2 N}~.
\ee 
The result still shows strong suppression in $N$.  The decision to single 
out $\sqrt{N}$ soft modes for enhancement is not essential.  We could 
declare any fraction of the long wave length modes which vanishes in 
the $N \to \infty$ limit as ``soft''.  A similar interpolation between 
the limits of uncorrelated and $\chi$RMT modes will always lead to a 
value of $\langle s^2 \rangle -1$ which vanishes as $N \to \infty$.  In short, the 
effects of a Thouless energy are expected to have a negligible effect 
on $\langle \la_i^2 \rangle$ over the entire microscopic spectrum.

The question is now how we can evaluate the effect on the zeros  from the
enhanced long wave length modes. To this end we introduce the distribution function of the fluctuations 
\be
\label{Ps}
P_N(s)\equiv \frac{1}{{\cal N}} s^\gamma e^{-\frac{s^2}{2\sigma^2}}
\ee
and fix $\gamma$ and $\sigma$ by the value of $\langle s^2 \rangle$
found above. ${\cal N}$ is the normalization.
In the RMT case we have $\gamma=N/2-1$ and $\sigma^2=2(N+1)/N^2$, 
while in the case where the long wavelength modes are enhanced we have 
$\gamma=N/\log(N)-1$ and $\sigma^2=\log(N)(1+\log(N)/(2N))/N$. 

The effect of the first $\sqrt{N}$ normal modes is different for large and small eigenvalues. Whereas for the smallest $\sqrt{N}$ eigenvalues it amounts to a breathing it means incoherent fluctuations for the larger ones. We now 
evaluate the effect of this breathing on the microscopic zeros.
Recall that the partition function for $N_f=1$ and $\nu=0$ is the 
average of the fermionic determinant with respect to the joint probability distribution, and that
$\prod_{i=1}^N(m^2-\la_i^2) = L_N^{-1/2}(N m^2)$. In the microscopic limit where
the quantity $\mu\equiv 2\,N\,m$ is fixed we have 
\be
L_N^{-1/2}\left(\frac{\m^2}{4Ns^2}\right) = \frac{(-1)^N}{N! 2^{2 N}} H_{2
  N}\left(\frac{\m}{2\sqrt{N}s}\right)\longrightarrow\frac{1}{\sqrt{N
    \pi}}\cos\left(\frac{\m}{s}\right)~. 
\ee
To investigate the correction to the microscopic partition function zeros
under the influence of the $\sqrt{N}$ longest wavelength normal modes we 
thus have to consider the following integral 
\be\label{Zm}
Z(\m) \sim
\frac{1}{{\cal N}\mathcal{K}}
\int_{0}^{\infty}ds\;\exp\left(\gamma\,\log\,s-\frac{s^2}{2\sigma^2}\right)\; 
\cos\left(\frac{\m}{s}\right)\;s^{2\sqrt{N}}~,  
\label{RZ}
\ee
where the normalization factor $\mathcal{K}$ is
\be
\mathcal{K} = \int_{0}^{\infty}ds\;P_N(s)\;s^{2\sqrt{N}}~. 
\ee
The factor $s^{2\sqrt{N}}$ comes from the rescaling in the fermionic determinant. 
For the  evaluation of (\ref{RZ}) we make use of the series expansion of the cosine 
\be
\cos \left(\frac{x}{s}\right) = \cos (x) + x\,\sin (x)\,\left( s-1 \right) + {\mathcal{O}(s-1)}^2~.
\ee
This yields 
\be
Z(\m) \sim
\cos (\m) + \m \sin (\m)\,\frac{\sqrt{2}\,\sigma\,\Gamma\left((\gamma+2\sqrt{N}+2)/2\right)}
{\Gamma\left((\gamma+2\sqrt{N}+1)/2\right)}+ \ldots ~.
\ee
Since $\gamma\gg\sqrt{N}$, we can use that \cite{AS}
\be
\frac{\Gamma(z+a)}{\Gamma(z+b)} \sim z^{a-b}\left[1 +
  \frac{(a-b)(a+b-1)}{2\,z} + \mathcal{O}\left(\frac{1}{z^2}\right)\right]~,
\ee
and finally find that in the case where we consider $\sqrt{N}$ RMT long wavelength modes the correction term to $\cos (\m)$ is of order $1/\sqrt{N}$, and in the case of $\sqrt{N}$ enhanced wavelength modes it is of order $\log{N}/\sqrt{N}$. In both cases the zeros of the partition function are unaffected in the microscopic limit.  

These results suggest a new kind of ``universality'' of the
microscopic partition function. As a specific example this universality shows 
that the microscopic zeros are unaffected even if the $\sqrt{N}$ longest
wavelength normal modes are enhanced in such a way that they interpolate
between Poissonian and RMT statistics.

\section{The microscopic limit}
\label{sect:micro}

So far we have been discussing the joint distribution function, the zeros 
of the partition function, and the $N$ positions that specify the collective maximum. Here we link this to the more familiar microscopic-eigenvalue
density. For a finite $N$ the eigenvalue density is found from the joint
distribution function as 
\be
\rho_N^{N_f,\nu}\;(\la, \{m_{f}\})  =
\int_{-\infty}^{\infty}\ldots\int_{-\infty}^{\infty} d\la_{2}\ldots
d\la_{N} P_N^{N_{f}, \nu}(\la,\la_2,\la_3,\ldots,\la_N; \{m_f\})~.
\ee
The double-microscopic spectral density is then defined
as \cite{SV, V, DN}
\be\label{micro}
\rho^{N_f,\nu}_s(\zeta; \mu_1,\ldots,
\mu_{N_f})\equiv\lim_{N\to\infty}\frac{1}{N}\rho_N^{N_f,\nu}\left(
\frac{\zeta}{N},\frac{\mu_1}{N},\ldots,
\frac{\mu_{N_f}}{N}\right),
\ee
and similarly for all other spectral correlations. The functional form of
the microscopic eigenvalue density has been derived in \cite{DN} and for
$N_f=1$  and topological charge $\nu$ it reads
\be\label{micro1}
\rho_s^{N_f=1,\nu}(\zeta; \mu) = \frac{|\zeta|}{2}\Big( J_{\nu-1}(\zeta)^2 
+ J_{\nu-2}(\zeta)J_{\nu}(\zeta) \Big) - |\zeta|\frac{J_{\nu-2}(\zeta)[\mu I_\nu(\mu)J_{\nu-2}(\zeta) 
+ \zeta I_{\nu-2}(\mu)J_{\nu}(\zeta)]}{(\zeta^2 + \mu^2)I_{\nu-2}(\mu)}~.
\ee
The partition function in the microscopic limit for one flavor is
proportional to $I_{\nu}(\mu)$.
The extrema of (\ref{micro1}) are given by 
\be\label{deriv}
\frac{\partial \rho^{(N_f=1,\nu)}_s(\zeta ;\mu)}{\partial\zeta} = 0 ~ ,
\ee
and are obviously functions of $\mu$. If the derivative in (\ref{deriv}) is
evaluated at $\zeta = \mu$ then the solutions of (\ref{deriv}) corresponding
to the maxima of $\rho_s^{(N_f=1,\nu)}$  coincide with the zeros of the $N_{f}=1$
microscopic partition function, $Z^{N_f=1,\nu}(m)$, for an imaginary
argument. To conclude: The peaks of the microscopic eigenvalue density are in 
one to one correspondence with the zeros of the partition function and by the 
trapping proven in section \ref{sect:trap} also to the positions which maximize the joint probability distribution.

\section{Conclusions and outlook}
\label{sect:conc}

In the present paper we have established an intimate relationship between
zeros of the partition function and the spectral properties of the Dirac
operator. The relation is derived within chiral random matrix theory and
applies to QCD Dirac spectra and partition function zeros near the origin (in 
the microscopic regime). Through the introduction of spectral normal modes we 
have tested the validity of the relationship when a finite Thouless energy is 
introduced. The observed independence of the microscopic zeros compliments
the existing universality studies. The present study treats one
flavour. For more flavours with degenerate masses, it is known that the zeros
of the microscopic partition function are not confined to the imaginary
axis \cite{KR}. While this makes the relation between the eigenvalues and the
zeros somewhat less direct, there is no reason to expect that the normal mode
analyis should not apply for any number of flavours.

We remark that the normal mode analysis is generically applicable and not 
a special feature of random matrix theory. In particular, the normal mode
analysis lends itself to a study of  the spectral properties of the Dirac
operator in lattice QCD. Such a study would be truly interesting in that it
would shed new light on the role of random matrix like correlations in
lattice gauge theories. Almost all of the normal modes are expected to be 
given by random matrix theory while only the very long wavelength modes
are determined by the detailed dynamics.

In a broader perspective this study may also be seen as the first step towards the establishment of a one-to-one correspondence between the zeros and the most likely eigenvalue  positions whenever the short range spectral correlations are random matrix like. If such a general relation were established then it could be used to argue that the critical exponents for the Yang-Lee edge in QCD must coincide
with the ones for the gap in the spectral density of the Dirac
operator. Such relationship, if true, would allow for substantial
simplifications  when trying to determine critical exponents by lattice
techniques.

\noi
{\sc Acknowledgments:}\\The authors wants to thank Poul Henrik Damgaard and
Thomas Wilke for useful discussions.

\end{document}